\documentclass[12pt]{article}

\usepackage{graphicx}
\usepackage{epsf}

\usepackage{amsmath,amssymb,float} 

\usepackage{pstricks}
\usepackage{color}

\def\beq{\begin{equation}}
\def\eq{\end{equation}}
\def\eeq{\end{equation}}

\def\centeron#1#2{{\setbox0=\hbox{#1}\setbox1=\hbox{#2}\ifdim
\wd1>\wd0\kern.5\wd1\kern-.5\wd0\fi
\copy0\kern-.5\wd0\kern-.5\wd1\copy1\ifdim\wd0>\wd1
\kern.5\wd0\kern-.5\wd1\fi}}
\def\ltap{\;\centeron{\raise.35ex\hbox{$<$}}{\lower.65ex\hbox{$\sim$}}\;}
\def\gtap{\;\centeron{\raise.35ex\hbox{$>$}}{\lower.65ex\hbox{$\sim$}}\;}

\def\lsim{\mathrel{\ltap}}

\def\MET{{\not \!  \! E}_T}

\def\chii0{\chi_i^0}
\def\chij0{\chi_j^0}

\def\GTLT{ \mathop{}_{<}^{>} }

\def\foursqr#1#2{{\vcenter{\vbox{
 \hrule height.#2pt
 \hbox{\vrule width.#2pt height#1pt \kern#1pt
 \vrule width.#2pt}
 \hrule height.#2pt
 \hrule height.#2pt
 \hbox{\vrule width.#2pt height#1pt \kern#1pt
 \vrule width.#2pt}
 \hrule height.#2pt
     \hrule height.#2pt
 \hbox{\vrule width.#2pt height#1pt \kern#1pt
 \vrule width.#2pt}
 \hrule height.#2pt
     \hrule height.#2pt
 \hbox{\vrule width.#2pt height#1pt \kern#1pt
 \vrule width.#2pt}
 \hrule height.#2pt}}}}
\def\psqr#1#2{{\vcenter{\vbox{\hrule height.#2pt
 \hbox{\vrule width.#2pt height#1pt \kern#1pt
 \vrule width.#2pt}
 \hrule height.#2pt \hrule height.#2pt
 \hbox{\vrule width.#2pt height#1pt \kern#1pt
 \vrule width.#2pt}
 \hrule height.#2pt}}}}
\def\sqr#1#2{{\vcenter{\vbox{\hrule height.#2pt
 \hbox{\vrule width.#2pt height#1pt \kern#1pt
 \vrule width.#2pt}
 \hrule height.#2pt}}}}

\def\figin{\epsfcheck\figin}\def\figins{\epsfcheck\figins}
\def\epsfcheck{\ifx\epsfbox\UnDeFiNeD
\message{(NO epsf.tex, FIGURES WILL BE IGNORED)}
\gdef\figin##1{\vskip2in}\gdef\figins##1{\hskip.5in}
\else\message{(FIGURES WILL BE INCLUDED)}%
\gdef\figin##1{##1}\gdef\figins##1{##1}\fi}
\def\DefWarn#1{}
\def\figinsert{\goodbreak\midinsert}
\def\ifig#1#2#3{\DefWarn#1\xdef#1{fig.~\the\figno}
\writedef{#1\leftbracket fig.\noexpand~\the\figno}%
\figinsert\figin{\centerline{#3}}\medskip\centerline{\vbox{\baselineskip12pt
\advance\hsize by -1truein\noindent\footnotefont{\bf
Fig.~\the\figno:\ } \it#2}}
\bigskip\endinsert\global\advance\figno by1}

\def\fig#1#2#3#4{\vskip 0.5cm \begingroup \midinsert \centerline{
\psfig{file=#1,width=#2}} \vskip 0.4cm
\global\advance\figno by 1
\centerline{\vbox{\baselineskip=12pt \noindent Figure \the\figno: #3}}
\endinsert \endgroup {\xdef#4{\the\figno}} }

\def\figcrop#1#2#3#4#5#6#7#8{\vskip 0.5cm \begingroup \midinsert \centerline{
\psfig{file=#1,width=#2,bbllx=#3,bblly=#4,bburx=#5,bbury=#6}} \vskip 0.4cm
\global\advance\figno by 1
\centerline{\vbox{\baselineskip=12pt \noindent Figure \the\figno: #7}}
\endinsert \endgroup {\xdef#8{\the\figno}} \vskip .5cm}

\def\figlabel#1{\xdef#1{\the\figno}}
\def\encadremath#1{\vbox{\hrule\hbox{\vrule\kern8pt\vbox{\kern8pt
\hbox{$\displaystyle #1$}\kern8pt}
\kern8pt\vrule}\hrule}}
\def\underarrow#1{\vbox{\ialign{##\crcr$\hfil\displaystyle
 {#1}\hfil$\crcr\noalign{\kern1pt\nointerlineskip}$\longrightarrow$\crcr}}}

\begin{document}

\begin{titlepage}

\begin{center}
\vspace*{-1cm}

\hfill RU-NHETC-2012-14 \\
\vskip 1.5in
{\LARGE \bf Searching for $t \to c h$ with Multi-Leptons} \\
\vspace{.15in}

\vskip 0.35in
{\large Nathaniel Craig},$^{1,2}$~
{\large Jared A. Evans},$^{1}$~
{\large Richard Gray},$^1$
\vskip0.1in
{\large  Michael Park},$^1$~
{\large Sunil Somalwar},$^1$~
{\large Scott Thomas},$^1$~
{\large Matthew Walker}$^1$

\vskip 0.25in
$^1${\em 
Department of Physics \\
Rutgers University \\
Piscataway, NJ 08854}

\vskip 0.12in
$^2${\em 
School of Natural Sciences
\\ Institute for Advanced Study \\
Princeton, NJ 08540 \\}

\vskip 0.4in

\end{center}

\baselineskip=16pt

\begin{abstract}

\noindent


The results of a
multi-lepton search conducted by the CMS collaboration 
with 5 fb$^{-1}$ of data collected from 7 TeV $pp$ collisions 
are used 
to place the first bound on the rare flavor-changing decay of the 
top quark 
to a Higgs boson and charm quark. 
Combining results from a number of exclusive 
three- and four-lepton search channels 
yields an estimated upper  limit 
of ${\rm Br}(t \to ch) < 2.7 \%$ for a Higgs boson mass of 125 GeV. 
The sensitivity of future dedicated searches for $t \to c h$ could be  
improved by adding exclusive 
same sign di-lepton channels, as well as by sub-dividing channels 
based on $b$-quark tagging and partial kinematic top quark and Higgs boson tagging.
This bound may be interpreted more widely within a range of 
new physics processes that yield final states 
with a $W$-boson in association with a Higgs boson. 
For such processes with kinematics that are similar 
to top--anti-top production and decay, the estimated limit 
on cross section times branching ratio corresponds to 
roughly $ \sigmaÊ\! \cdot  \! {\rm Br}( pp \to  WhX ) < 9$ pb.

\end{abstract}

\end{titlepage}

\baselineskip=17pt

\newpage


\section{Introduction}

Once the existence of a Standard Model-like Higgs boson is established
at the Large Hadron Collider (LHC) 
it will be
possible to search for new physics 
produced in association with the Higgs. 
The myriad decay modes of the Higgs boson 
offer a wide range 
of possibilities for such searches. 
One attractive possibility is to focus on the leptonic final states of the Higgs boson 
in association with additional leptons coming from other particles accompanying 
the Higgs.   
Multi-lepton signatures originating from Standard Model production and decay of the
Higgs boson itself
provide considerable sensitivity \cite{ContrerasCampana:2011aa}, 
and in conjunction with additional leptons could 
provide powerful probes of 
non-Standard Model processes that include a Higgs.

One class of non-Standard Model 
processes of interest are those in which the Higgs boson 
appears only rarely in association with other particles.   
In this case, observation of a new physics 
process requires a large production cross section, 
making it fruitful to consider Standard Model processes 
with large production cross section.   
The production of top--anti-top quark pairs 
is particularly attractive in this 
respect, with a cross section of 100's of pb
at the LHC.  
This suggests looking for Higgs bosons in the decay products of the top quark, 
such as would arise through the rare neutral 
flavor-changing transition to a charm quark, $t \to c h.$ 
In fact, no symmetry forbids this decay, 
but the Standard Model contribution to the branching ratio 
suffers both GIM and second-third generation mixing suppression and is 
extremely small, of order 
${\rm Br}(t \to ch)_{\rm SM} \simeq 10^{-13}-10^{-15}$ 
\cite{Eilam:1990zc, Mele:1998ag, AguilarSaavedra:2004wm}. 
Thus a positive observation of the process $t \to c h$ 
well above the Standard Model rate would be a convincing indication 
of new physics beyond the Standard Model. 

Pair production of top--anti-top 
followed by the rare decay $t \to c h$ 
gives rise to multi-lepton final states with up to 
five leptons.  
The leading processes involve leptonic charged-current 
decay of one of the top quarks, 
$t \to Wb$ with $W \to \ell \nu$, and flavor-changing decay 
of the other top quark, $\bar{t} \to \bar{c} h$ with 
leptonic final state decay modes of the Higgs boson. 
These include
$h \to WW^* \to \ell \nu \ell \nu$, and 
$h \to \tau \tau$ with leptonic decay of the 
tau-leptons, $\tau \to \ell X$, 
as well as 
$h \to ZZ^* \to jj \ell \ell, \nu \nu \ell \ell, \ell \ell \ell \ell$. 
Hadronic decay of one of the top quarks, 
$t \to Wb$ with $W \to jj$, and flavor-changing decay of the other top quark,
$\bar{t} \to \bar{c} h$ with $h \to Z Z^* \to  \ell \ell \ell \ell$ also contributes. 
Such multi-lepton final states have relatively low Standard Model backgrounds, 
making them promising targets for a multi-lepton search. 

To investigate the utility of searching for 
$t \to ch$ in this way, we make use of the results of a multi-lepton 
search conducted by the CMS collaboration with 5 fb$^{-1}$ of 
data collected from 7 TeV $pp$ collisions  \cite{Chatrchyan:2012ye} 
to estimate a limit on the branching ratio ${\rm Br}(t \to ch)$. 
The power of this search lies in the combination of numerous exclusive channels. 
While any individual channel alone is not necessarily 
significant, the exclusive combination across multiple channels is 
found to provide an interesting sensitivity to ${\rm Br}(t \to c h)$ 
at the percent-level.   
To our knowledge this is the first use of the Higgs boson as a probe for new physics 
in existing data.  

The neutral flavor-changing decay of a top quark to the Higgs boson and charm quark, 
$t \to c h$, is 
of interest because 
it provides a 
direct probe of flavor violating couplings 
to the Higgs sector for the quark that is most strongly coupled to that sector. 
Previous probes of 
flavor violating couplings to the Higgs sector 
for the lighter quarks have  
been only indirect. 
It is also of general interest because 
up-type quark flavor violation is less well constrained 
than that for down-type quarks. 
Given that this process has not been investigated experimentally 
at any level previously, 
the percent-level bound on ${\rm Br}(t \to ch)$ obtained 
here begins to open up an interesting
new window into flavor violating physics.

In \S 2 we present an effective operator 
analysis of the rare decay 
$t \to ch$ and give the relation between the branching ratio and 
new physics scale of the leading operator that contributes to this
process.  
In \S 3 we review multi-channel multi-lepton searches 
and compare the results of a CMS search 
with our simulation of top quark pair production 
and decay, including $t \to c h$, to 
obtain the first limits on ${\rm Br}(t \to c h)$ for a 
Standard Model-like Higgs boson.
We also suggest improvements that could increase the intrinsic sensitivity 
of future dedicated multi-lepton searches for $t \to ch$.
In \S 4 we discuss the wider applicability of this result 
couched in terms of a cross section times branching limit 
on new physics that yields  final states with a 
$W$-boson in association with a Higgs boson.

It should be noted that, although 
throughout we refer to the flavor-violating decay of the top quark 
to a Higgs boson as $t \to c h$, since the identity of the 
charm quark is not 
integral to the analysis, the discussion and results apply 
more generally to the decay $t \to Xh$ with 
inclusive $X$ final states.


\section{Effective Operator Description of $t \to c h$}

New physics contributions to the flavor-violating top quark decay 
$t \to c h$ may be encoded in an effective field theory description 
of the operators that can contribute to this 
process. 
For the field content of the minimal Standard Model, 
the leading coupling of the Higgs boson to up-type 
quarks is through the renormalizable dimension-four Yukawa coupling 
\beq
\lambda_{ij} Q_i H \bar u_j + {\rm h.c.} 
\label{yukawa} 
\eq
where the quark fields are two component complex 
Weyl Fermions.  
The most relevant sub-leading interactions 
coupling the Higgs to up-type quarks 
come from dimension-six operators. 
Up to operator relations at this order, these may written in terms of a single 
non-renormalizable operator 
\beq
\frac{\xi_{ij}}{M^2} H^\dag H Q_i H \bar u_j + {\rm h.c.} 
\label{eq:effop}
\eq
At this order in an effective field theory expansion, both 
the Yukawa coupling (\ref{yukawa}) 
and dimension-six operator (\ref{eq:effop})
contribute to the up-quark mass matrix and effective 
coupling to the physical Higgs boson 
\beq
m_{ij} u_i \bar{u}_j + \lambda_{ij}^h h u_i \bar u_j + {\rm h.c.}
\label{mhcouplings}
\eq
where the up-quark mass matrix is given by 
\begin{equation}
m_{ij} = \frac{v}{\sqrt{2}} \left[ \lambda_{ij} + \frac{v^2}{2M^2} \xi_{ij} \right] 
\equiv \frac{v}{\sqrt{2}} \lambda_{ij}^m
\end{equation}
and where $H^0 = \frac{1}{\sqrt{2}}(v+h)$ and 
$\lambda_{ij}^m$ is the mass effective Yukawa coupling.
The Higgs effective Yukawa coupling, 
$\lambda_{ij}^h$,  of the physical Higgs boson 
to up-quarks 
is given at this order in the effective field theory description 
by the derivative of the mass matrix with respect to the Higgs 
expectation value 
\begin{equation}
\lambda_{ij}^h = \frac{\partial m_{ij}}{\partial v} = 
\frac{1}{\sqrt{2}} \left[ \lambda_{ij}^m + \frac{v^2}{M^2} \xi_{ij} \right]
\label{hcouplings} 
\end{equation}
Since the mass effective Yukawa is by definition diagonal in the 
mass basis, flavor-violating interactions come only from the 
second term in parentheses in (\ref{hcouplings}). 
Misalignment between the mass and Higgs effective Yukawa couplings, 
$\lambda^m_{ij}$ and $\lambda^h_{ij}$,
vanishes in the $M \to \infty$ limit. 

The partial
decay width of the top quark to a Higgs boson and massless charm quark 
from the effective Higgs interaction 
(\ref{mhcouplings})
with flavor violating couplings (\ref{hcouplings})
is given by
\begin{equation}
\Gamma(t \to ch) = \frac{( |\xi_{tc}|^2 + |\xi_{ct}|^2) m_t}{128 \pi G_F^2 M^4} \left( 1 - \frac{m_h^2}{m_t^2} \right)^2
\end{equation}
where $G_F^{-1} = \sqrt{2} v^2$. 
For comparison, the partial decay width of the top quark to the $W$-boson and massless $b$-quark
through the minimal charged current interaction 
is 
\begin{equation}
\Gamma(t \to Wb) = \frac{G_F m_t^3 |V_{tb}|^2}{8 \pi \sqrt{2}} \left( 1 - \frac{m_W^2}{m_t^2} \right)^2 \left(1 + \frac{2 m_W^2}{m_t^2} \right)
\end{equation}
Assuming ${\rm Br}(t \to Wb)$ is close to unity, the
leading order branching ratio for $t \to ch$ is then given by
\begin{equation} \label{eq:br}
{\rm Br}(t \to ch) \simeq 
\frac{ |\xi_{tc}|^2 +|\xi_{ct}|^2  }{8 \sqrt{2} G_F^3 m_t^2 M^4 |V_{tb}|^2} 
\frac{(1 - m_h^2/m_t^2)^2}{(1-m_W^2/m_t^2)^2(1+2 m_W^2/m_t^2)}
\end{equation}
For Higgs boson and top quark masses of 
$m_h = 125$ GeV and 
$m_t = 173.5$ GeV respectively, 
the numerical value of the branching ratio in terms of the
dimension-six operator scale and flavor-violating 
Higgs effective Yukawa coupling are
\beq
{\rm Br}(t \to ch) \simeq \left( {150{\rm \; GeV} \over M / 
\sqrt[4]{ |\xi_{tc}|^2 +|\xi_{ct}|^2}  }  \right)^4
\simeq 0.29 ~ \left( |\lambda^h_{tc}|^2 + |\lambda^h_{ct}|^2 \right) 
\eq


\section{A multi-lepton search for $t \to c h$} \label{sec:search}

Multilepton searches at hadron colliders 
provide great sensitivity to new physics 
processes.  
In this work we follow and use the results of the multi-lepton search strategies adopted by the 
CMS collaboration \cite{Chatrchyan:2012ye,Chatrchyan:2011ff}. 
The sensitivity to new physics arises from dividing three- or more-lepton final states into 
a large number of exclusive 
search channels based on 
lepton flavor and charge combinations, 
hadronic activity, missing transverse energy, 
and the kinematic properties of the leptons in an event. 
We first review the details of this search strategy before applying it to obtain a bound 
on ${\rm Br}(t \to ch)$.

\subsection{Multi-lepton signal channels}\label{subsec:channels}

Standard Model backgrounds to multi-lepton searches 
for new physics are small and may be further reduced by imposing 
cuts on hadronic activity or missing energy. In this case hadronic 
activity is characterized by the variable $H_T$, the scalar sum of 
the transverse jet energies for all jets passing the preselection cuts. 
The missing transverse energy, MET, is given by the magnitude of the 
vector sum of the momenta of all reconstructed 
objects. 
Both $H_T$ and MET are sensitive discriminating observables for new physics
in a given lepton flavor and charge channel.  

The CMS multilepton search \cite{Chatrchyan:2012ye} 
exploits the background discrimination of $H_T$ and MET in the following way: Events with $H_T > 200$ (MET $>50$) GeV are assigned 
HIGH $H_T$ (MET), while those with $H_T < 200$ (MET $< 50$) GeV 
are assigned LOW $H_T$ (MET). 
The HIGH $H_T$ and HIGH MET requirements (individually or in combination) lead to a significant reduction in Standard Model backgrounds.\footnote{It is also possible to reduce backgrounds using an $S_T$ variable defined to be the scalar sum of 
MET, $H_T$, and leptonic $p_T$  \cite{Chatrchyan:2012ye}, but for simplicity we will not make use of $S_T$ here.} 

Further background reduction may be accomplished with a $Z$-boson veto, 
in which the invariant mass of opposite-sign same-flavor (OSSF)
lepton pairs is required to lie outside a $75-105$ GeV window 
around the $Z$ mass; we simply denote events passing the $Z$ veto as No $Z$. 
In the case of $3 \ell$ events, it is also useful to differentiate between events with no 
OSSF pairs, 
which we denote DY0 to indicate 
no possible Drell-Yan pairs, and one OSSF pair which we denote DY1. 
Although the CMS multi-lepton analysis \cite{Chatrchyan:2012ye,Chatrchyan:2011ff}
also includes channels with one or more objects consistent with hadronically 
decaying $\tau$-leptons, 
in this analysis we will focus our attention on $\ell = e, \mu$ only. 
We do implicitly include leptonically decaying $\tau$-leptons in our analysis, 
which for all practical purposes in the detector are simply $e$- or $\mu$-leptons. 

The $3 \ell$ or $4 \ell$ channels may be divided into 20 possible 
combinations of $H_T$ HIGH/LOW; MET HIGH/LOW; $Z$/No $Z$; and DY0/DY1. 
The 20 channels are presented in Table~\ref{tab:SM}. For each of the $3 \ell$ and $4 \ell$ categories, channels are listed from top to bottom in 
approximately 
descending order of backgrounds, or equivalently 
ascending order of sensitivity, with the last such channel at the bottom 
dominated by Standard Model backgrounds. However, all channels contribute to the limit.

\begin{table}
\begin{center}
{\small
\begin{tabular}{lllccc}
\hline \hline \\

& & &   Observed & Expected & Signal  \\

   4 Leptons & \\ \\
~~MET HIGH & HT HIGH & No Z   & 0 & 0.018 $\pm$ 0.005 & 0.02  \\
~~MET HIGH & HT HIGH &~~~~~Z        & 0 & 0.22 $\pm$ 0.05 & 0.0 \\
~~MET HIGH & HT LOW & No Z   & 1 & 0.2 $\pm$ 0.07 & 0.11 \\
~~MET HIGH & HT LOW &~~~~~Z         & 1 & 0.79 $\pm$ 0.21  & 0.04 \\
~~MET LOW & HT HIGH & No Z    & 0  & 0.006 $\pm$ 0001 & 0.0 \\
~~MET LOW & HT HIGH &~~~~~Z         & 1& 0.83 $\pm$ 0.33 & 0.04 \\
~~MET LOW & HT LOW & No Z     & 1 & 2.6 $\pm$ 1.1 & 0.08 \\
~~MET LOW & HT LOW &~~~~~Z          & 33& 37 $\pm$ 15 & 0.15 \\

  & & \\
  3 Leptons &\\  \\

~~MET HIGH & HT HIGH & DY0              & 2 & 1.5 $\pm$ 0.5 & 0.48 \\
~~MET HIGH & HT LOW & DY0               & 7 & 6.6 $\pm$ 2.3  & 2.1 \\
~~MET LOW & HT HIGH & DY0                & 1 & 1.2 $\pm$ 0.7  & 0.26 \\
~~MET LOW & HT LOW & DY0                & 14 & 11.7 $\pm$ 3.6 & 1.68 \\
~~MET HIGH & HT HIGH & DY1 No Z      & 8 & 5 $\pm$ 1.3 & 1.54  \\
~~MET HIGH & HT HIGH & DY1~~~~~~Z   & 20 & 18.9 $\pm$ 6.4 & 0.41 \\
~~MET HIGH & HT LOW & DY1 No Z       & 30 & 27 $\pm$ 7.6  & 5.8 \\
~~MET HIGH & HT LOW & DY1~~~~~~Z    & 141 &  134 $\pm$ 50 & 2.0 \\
~~MET LOW & HT HIGH & DY1 No Z       & 11 & 4.5 $\pm$ 1.5 & 0.80 \\
~~MET LOW & HT HIGH & DY1~~~~~~Z   & 15 & 19.2 $\pm$ 4.8 & 0.72  \\
~~MET LOW & HT LOW & DY1 No Z          &  123  & 144 $\pm$ 36  & 3.1 \\
~~MET LOW & HT LOW & DY1~~~~~~Z    & 657 & 764 $\pm$ 183 & 2.4  \\


   & &  \\
\hline \hline
\end{tabular}}
\caption{
Observed number of events, expected number of background events, and expected  number
of $t \to c h$ signal events with 
${\rm Br}(t \to c h) = 1\%$ in various CMS 
multi-lepton channels after acceptance and efficiency for 5 fb$^{-1}$ of 7 TeV proton-proton collisions.
HIGH and LOW for MET and HT indicate $\MET \GTLT $ 50 GeV and $H_T \GTLT  200$ GeV respectively. 
DY0 $\equiv \ell^{\prime \pm} \ell^{\mp} \ell^{\mp}$, DY1 $\equiv \ell^{\pm} \ell^+ \ell^-, \ell^{\prime \pm} \ell^+ \ell^- $, 
for $\ell = e, \mu$. 
No Z and Z indicate $|m_{\ell \ell} - m_Z| \GTLT 15$ GeV for any opposite sign same flavor pair. 
}
\label{tab:SM}
\end{center}

\end{table}

\subsection{Simulation details}

We closely follow the CMS multilepton analysis \cite{Chatrchyan:2012ye}, 
applying the same cuts to our signal sample and making use of the 
CMS background estimates and observations with 5 fb$^{-1}$ of 7 TeV $pp$ collision data. 
For our signal, we simulate $t \bar t$ production events 
with one side decaying through 
conventional charged current interaction 
via $t \to W b$ and the other side decaying via $t \to c h$.  
For definiteness we take $m_h = 125$ GeV with Standard Model branching ratios. 
For simulating signal processes, we have used MadGraph v4 \cite{Maltoni:2002qb,Alwall:2007st} and rescaled the $t \bar t$ production cross section to the NLO value 
$\sigma(pp \to {t \bar t}) = 165$ pb at 7 TeV \cite{Moch:2008ai}. The Higgs boson was decayed inclusively using BRIDGE \cite{Meade:2007js}. The branching ratios and total width for Higgs decay in BRIDGE were taken from the LHC Higgs Cross Section Group \cite{LHCHiggsCrossSectionWorkingGroup:2011ti}. Subsequent showering and hadronization effects were simulated using Pythia \cite{Sjostrand:2006za}. Detector effects were simulated using PGS \cite{PGS} with the isolation algorithm for muons 
modified to more accurately reflect the procedure used by the CMS collaboration. 
In particular, we introduce a \texttt{trkiso} variable 
 for each muon \cite{Gray:2011us}. 
 The variable \texttt{trkiso} is defined to be the sum $p_T$ of all tracks, ECAL, and HCAL deposits within an annulus of inner radius 0.03 and outer radius 0.3 in $\Delta R$ surrounding a given muon. 
Isolation requires that for each muon, \texttt{trkiso}/$p_{T_\mu}$ is 
less than 0.15. 
The efficiencies of PGS detector effects were normalized by simulating the TeV3 mSUGRA benchmark studied in \cite{Chatrchyan:2011ff} and comparing the signal in 3$\ell$ and $4 \ell$ channels. To match efficiencies with the CMS study we applied an efficiency correction of 0.87 per lepton to our signal events
\cite{ContrerasCampana:2011aa}. 
We applied preselection and analysis cuts  in accordance with those used
in the CMS analysis \cite{Chatrchyan:2012ye}.
A total of 500,000 events were simulated to give good statistical 
coverage of all the relevant multi-lepton channels.


\subsection{Results}

The multi-lepton final states coming from $t \to c h$ in $t \bar t$ pair production 
arise mainly from 
charged current decay of one top quark, 
$t \to W b$ with $W \to \ell \nu$, and 
 flavor-violating decay of the other top quark, 
$\bar{t} \to \bar{c} h$, with $h$ decaying to final states with two or more leptons.
The most relevant Higgs final states are those with two leptons that arise from 
$h \to WW^* \to \ell \nu \ell \nu$ and 
$h \to \tau \tau$ with leptonic decays of the tau-leptons, 
$\tau \to \ell X$, 
as well as 
$h \to ZZ^* \to jj \ell \ell, \nu \nu \ell \ell$.
All of these decay modes give three-lepton final states.  
Although the total branching ratio of the Higgs 
to two leptons is comparable for $h \to WW^*$ and $h \to ZZ^*$, 
leptons coming from $Z$ and/or $Z^*$ decays are less significant 
because they fall into higher-background DY1 channels with 
either $Z$ or No $Z$. 
In contrast, pairs of leptons coming from 
$WW^*$ decay are uncorrelated in flavor and 
fall into lower-background DY0 channels, in addition to 
the higher-background DY1 channels. 
There are additionally four- and five-lepton final states from 
charged current decay of one top quark, 
$t \to W b$ with $W \to jj$ or $\ell \nu$ respectively, 
and  
flavor-violating decay of the other top quark, 
$\bar{t} \to \bar{c} h$, with $h \to ZZ^* \to \ell \ell \ell \ell$. 
The small total branching ratio for these final states 
makes them less significant than the three-lepton 
final states in obtaining a bound from the 5 fb$^{-1}$
of integrated luminosity in the CMS search \cite{Chatrchyan:2012ye}. 
However, in the future with more integrated luminosity, these 
channels should contribute more significantly to the sensitivity for 
$t \to ch$. 

The signal contributions to each of the exclusive multi-lepton channels 
are shown in Table 1. 
Events are entered in the table exclusive-hierarchically from the top to the bottom.
In this way each event appears only once in the table, and in the lowest 
possible background channel 
consistent with its characteristics.  
The strongest limit-setting channels 
for $t \to ch$ 
are those with three leptons. 
The best limits come from [MET HIGH, HT LOW, DY1 No Z], 
which alone constrains ${\rm Br}(t \to c h) <  3.7\%$, and  
[MET HIGH, HT LOW, DY0], which constrains ${\rm Br}(t \to c h) < 4.2 \%$. 
In each case the lack of a reconstructed $Z$ or OSSF lepton pair reflects 
the contributions from $h \to WW^*$ and $h \to \tau \tau$, while the 
MET comes predominantly from neutrinos emitted in the $W$ and $\tau$-lepton 
decays. 
Significant limits also come from the channel [MET HIGH, HT HIGH, DY1 No Z], 
which constrains ${\rm Br}(t \to c h) <  6.5 \%$; and  [MET  LOW, HT LOW, DY0], 
which constrains ${\rm Br}(t \to c h) <  7.9 \%$; these likewise reflect dominant 
contributions from the Higgs decays 
$h \to WW^*$ and $h \to \tau \tau$. All other channels give constraints on 
the branching ratio that are weaker than $10 \%$ in an individual channel.

Although limits may be placed on the signal from 
any individual channel in the multi-lepton search, 
the greatest sensitivity comes from combining all exclusive channels.
Combining all multilepton channels, we find that the 5 fb$^{-1}$ 
multi-lepton CMS results \cite{Chatrchyan:2012ye} yield an observed 
limit of ${\rm Br}(t \to ch) < 2.7 \%$, with an expected limit ${\rm Br}(t \to ch) < 1.7 \%$. 
This corresponds to a bound on the scale of the 
dimension-six effective operator (\ref{eq:effop}) 
introduced in \S 2 
of $M^2/ \sqrt{ |\xi_{tc}|^2 + |\xi_{ct}|^2} > (370 {\rm\; GeV})^2$ 
or equivalently on the flavor-violating Higgs Yukawa couplings 
(\ref{hcouplings}) of $\sqrt{|\lambda^h_{tc}|^2 + |\lambda^h_{ct}|^2}  < 0.31$.  
This limit represents a combined Bayesian 95\% CL limit computed 
using the observed event counts, background estimates, and systematic errors listed in Table~\ref{tab:SM}. 

An upper limit on the branching ratio 
${\rm Br}(t \to ch) $ can also be expressed in terms of a limit on the 
cross section times branching ratio 
$ \sigmaÊ\! \cdot \! {\rm Br}( pp \to t \bar{t} \to Wbhc )$. 
This is related to the cross section and branching ratio individually 
by 
$ \sigmaÊ\! \cdot \! {\rm Br}( pp \to t \bar{t} \to Wbhc ) \simeq 
    \sigma ( pp \to t \bar{t}) \cdot 2~{\rm Br}(t \to hc)$
where the factor of two accounts for combinatorics of the top 
quark decay. 
With this, our estimate for the observed upper limit of 
${\rm Br}(t \to ch) < 2.7 \%$
corresponds to 
$ \sigmaÊ\! \cdot \! {\rm Br}( pp \to t \bar{t} \to Wbhc ) < 8.9$ pb
for 7 TeV $pp$ collisions. 
While this limit is specific to the acceptance and efficiency 
associated to top--anti-top production and decay, 
it does give a rough indication of the cross section times 
branching limit that would be obtained from the results of 
the CMS multi-lepton search \cite{Chatrchyan:2012ye} for other new physics 
processes $pp \to WhX$ with similar kinematics. 

The sensitivity of future dedicated multi-lepton searches for flavor-changing 
top quark decay $t \to ch$ could 
be improved in a number of ways.  
The most straightforward improvement would be to include 
the CMS 
exclusive multi-lepton 
channels that contain $\tau$-leptons. 
For simplicity these were neglected in this study. 
These channels have higher backgrounds, but would contribute 
a bit to the overall sensitivity.  
Another improvement 
would be to sub-divide the exclusive multi-lepton channels 
according to whether there are tagged $b$-quarks in an event.  
The $t \to ch$ signal has both a $b$- and $c$-quark in the final state, 
and so would fall primarily in the $b$-tagged channels.  
Although there is background from $\bar{t} t$ production with fully-leptonic 
decay and a fake lepton in these channels, other Standard Model backgrounds 
from, e.g. $WZ$ production with fully leptonic decay, would be reduced 
in these channels.  
Yet another possibility would be to incorporate exclusive 
same-sign di-lepton 
channels, again with $b$-quark tagging sub-division \cite{Chatrchyan:2012sa}.
Although the backgrounds in these channels are by definition 
larger than those of three- or more-lepton channels, 
this would bring in other relevant final states of the $\bar{t} t$ signal 
such as charged current decay of 
one top quark, $t \to Wb$ with $W \to \ell \nu $, and 
flavor-violating decay of the other top quark, 
$\bar{t} \to \bar{c}h$ with $h \to WW^* \to \ell \nu jj$. 
Since the Higgs boson is neutral, the charges of the two leptons 
from these decays are 
uncorrelated and same-sign half the time.  
Finally, further signal specific sub-divisions of channels could 
be utilized based on partial kinematic tagging information of the 
top quark and/or Higgs boson to isolate 
regions of phase space that are populated only by the signal.  

We emphasize that in respect to possible improvements focused 
at the $t \to c h$ signal, 
the current work represents a proof of principle illustrating the power 
of the CMS exclusive channel multi-lepton search strategy 
\cite{Chatrchyan:2012ye,Chatrchyan:2011ff}
that may be 
extended for certain new physics signals 
by a targeted refinement of the search channels.



\section{Conclusions}\label{sec:conc}

The discovery of a Standard Model-like Higgs opens the door to a plethora of new searches that employ Higgs decay products to probe new physics processes that involve Higgs boson associated
production or decay. 
In this paper we have studied one of 
the simplest such processes, the rare flavor-violating top quark 
decay to a Higgs boson and charm quark, 
$t \to c h$. 
Using the results of the CMS multi-lepton search with 
5 fb$^{-1}$ of 7 TeV $pp$ 
collision data \cite{Chatrchyan:2012ye}, 
we obtain the estimated upper bounds of ${\rm Br}(t \to ch) < 2.7 \%$ and 
$ \sigmaÊ \cdot  {\rm Br}( pp \to t \bar{t} \to Wbhc ) < 8.9$ pb
for a 125 GeV Standard Model Higgs boson with Standard Model 
branching ratios.  
Future multi-lepton searches at the LHC optimized for this signal, 
including $\tau$-lepton channels, exclusive same-sign di-lepton channels,
sub-division of channels based on $b$-quark tagging, 
and with increasing integrated luminosity,
should be able improve the sensitivity to $t \to c h$  considerably.   

The results presented here should be more widely applicable to 
a range of new physics processes that yield final states 
with a $W$-boson in association with a Higgs boson.  
For processes with kinematics that are similar to top--anti-top production 
and decay, 
our estimated 
bound from the CMS multi-lepton search \cite{Chatrchyan:2012ye}
corresponds very roughly to  
$ \sigmaÊ \cdot  {\rm Br}( pp \to  WhX ) \lsim 9$ pb. 
Just one example of many such new physics processes that are of interest 
is 
production of supersymmetric wino- or Higgsino like chargino and neutralino,
either directly or from cascade decays, 
with decay of the chargino to a $W$-boson and lighter 
neutralino or the Goldstino, and decay of the neutralino to a Higgs boson and 
a lighter neutralino or the Goldstino, 
$pp \to X \to \chi^{\pm} \chi^0_i Y \to Wh \chi^0_j \chi^0_j Y$. 
In many scenarios the branching ratios 
${\rm Br}(\chi^{\pm}  \to W \chi^0_j)$ and 
${\rm Br}(\chi^0_i \to h \chi^0_j)$ can approach unity \cite{Matchev:1999ft}. 
While the upper limit on the cross section times branching ratio 
obtained above does not quite bound direct 
electroweak chargino--neutralino production with these decays, 
it would provide bounds on certain scenarios 
with strong superpartner production
where the chargino and neutralino are emitted in cascade decays.  
The future improvements to exclusive channel 
multi-lepton searches mentioned above 
would improve the sensitivity also to these supersymmetric 
processes with associated Higgs bosons.  
In particular, direct chargino-neutralino production would yield 
final states without $b$-quarks, and so would appear 
as signal in the $b$-quark anti-tagged subdivision of exclusive 
same sign di-lepton and multi-lepton channels.  

The Higgs boson will provide a new calibration for experimental 
physics at high energy colliders. 
Higgs boson leptonic decay modes are  
but one of many possible applications of Higgs decays to the search for new physics.


\bigskip
\bigskip

{ \Large \bf Acknowledgments}

\smallskip \smallskip

\noindent
We thank Emmanuel Contreras-Campana
and Amit Lath for useful conversations. 
The research of NC, JE, MP and ST was supported in part by DOE grant DE-FG02-96ER40959. 
The research of RG, SS and MW was supported in part by NSF grant PHY-0969282. 
NC gratefully acknowledges the support of the Institute for Advanced Study.


\end{document}